\documentclass[final,twocolumn,twoside,10pt,a4paper]{IEEEtran} %,compsoc
\usepackage{graphicx,algorithm,algorithmic,multirow,flushend}
\usepackage{latexsym,color,float,amsmath,amssymb,subfigure}
\begin{document}

\title{Coordinating Interfering Transmissions in Cooperative Wireless LANs}

\author{Antonios Argyriou,~\IEEEmembership{Member,~IEEE}\thanks{Manuscript received November 21, 2010; revised April 1, 2011; June 16, 2011; accepted August 21, 2011. The associate editor coordinating the review of this paper
and approving it for publication was Xianbin Wang.}\thanks{A. Argyriou would like to acknowledge the support from the European Commission through the Marie Curie Intra-European Fellowship WINIE-273041 and the STREP project CONECT
(FP7ICT257616).}\thanks{A. Argyriou is with the Department of Computer and Communication Engineering, University of Thessaly, Greece (Email: anargyr@ieee.org).}}

\maketitle

\markboth{IEEE Transactions on Wireless Communications, Vol. XX, No. XX, XXXX 2011}{A. Argyriou: Coordinating Interfering Transmissions in Cooperative Wireless LANs}%

\graphicspath{{figures/}}%

\begin{abstract}
In this paper we present a cooperative medium access control (MAC) protocol that is designed for a physical layer that can decode interfering transmissions in distributed wireless networks. The proposed protocol pro-actively enforces two independent packet transmissions to interfere in a controlled and cooperative manner. The protocol ensures that when a node desires to transmit a unicast packet, regardless of the destination, it coordinates with minimal overhead with relay nodes in order to concurrently transmit over the wireless channel with a third node. The relay is responsible for allowing packets from the two selected nodes to interfere only when the desired packets can be decoded at the appropriate destinations and increase the sum-rate of the cooperative transmission. In case this is not feasible, classic cooperative or direct transmission is adopted. To enable distributed, uncoordinated, and adaptive operation of the protocol, a relay selection mechanism is introduced so that the optimal relay is selected dynamically and depending on the channel conditions. The most important advantage of the protocol is that interfering transmissions can originate from completely independent unicast transmissions from two senders. We present simulation results that validate the efficacy of our proposed scheme in terms of throughput and delay.
\end{abstract}


\section{Introduction}
\PARstart{O}{ne} of the most undesired side-effects of wireless communications systems is interference. In wireless networks, where several nodes share the medium, interference is avoided with mechanisms that orthogonalize transmissions. The classic examples of such mechanisms include frequency division multiple access (FDMA), time division multiple access (TDMA), code division multiple access (CDMA), and finally random access protocols like carrier sense multiple access with collision avoidance (CSMA/CA)~\cite{bertsekas:data-networks}. However, besides channel orthogonilization, there have been several additional techniques throughout the years that attempt to combat this effect~\cite{book:fundamental-wireless}. In more recent years, there is a trend to exploit interference in order to increase the network capacity~\cite{dankberg97,larsson05,popovski06b,zhang:physical-layer-nc,katabi07a}. This technique is usually referred to as analog network coding (ANC) and we can first identify it in~\cite{dankberg97}, although not with this term. With ANC network capacity is increased since concurrent interfering transmissions are allowed. Nodes listen to transmissions and then forward the unprocessed analog signals to destination nodes where various algorithms for interference cancelation can be applied in order to retrieve the signal of interest~\cite{katabi07a,argyriou:twc-ancol}. The removal of an interfering signal is possible with ANC when this signal is known at the receiver. A scenario where this might be the case is in multihop networks when the receiver had transmitted in the past the required signal in the form of a complete packet. By removing the previous assumption, we investigated the potential improvement of ANC in the sum-rate of a simple relay network with two completely independent senders/receivers and one relay in~\cite{argyriou:twc-ancol}. One of the main results was that if two packets, that originate from different senders and are directed towards different receivers, interfere partially or entirely in the time domain, the subsequent forwarding of the mixed packets can work in favor of both unicast transmissions by increasing the total sum-rate. In this paper we take this result and we attempt to utilize it in more practical networks where several nodes contend for the medium. We consider an extended and more realistic wireless ad hoc network where issues like channel estimation, medium access, and relay selection must be addressed.

In this paper we focus on the development of basic elements of a distributed cooperative random access MAC protocol that operates with an underlying physical layer (PHY) that employs ANC. It is important to stress at this point that we adopt the random access MAC principle due to its simplicity, ease of implementation in a distributed setting, and the widespread adoption in practical systems. Based on the previous choice and with the assumption that packet transmissions are allowed to interfere with a mechanism like ANC, we seek to identify the necessary algorithmic components that should be embedded a classic random access MAC protocol. Four specific algorithms of varying complexity are presented in this work. First, there is a need for a new channel access scheme that supports cooperative transmissions, next an algorithm for channel information exchange and estimation, rate estimation of the potential cooperative transmissions, and finally there is a need for an algorithm that disentangles and decodes the interfered signals. With these algorithms, the proposed cooperative MAC protocol fulfills first and foremost one basic task, that is it identifies when (and if) packets/signals can interfere. This task is performed by the relays in the wireless network that can act as "coding nodes" that subsequently forward the coded/interfered packets. Therefore, the relays in our protocol implement the bulk of the required intelligence in the sense that they make the decision whether a cooperative transmission with ANC is effective before it is allowed. The signal recovery algorithm needs only to be executed at the destinations as a final step in the overall transmission process.

\subsection{Related Works}
\label{sec:related-works}
\begin{figure}[t]%
\begin{center}
  \includegraphics[keepaspectratio,width = 0.5\linewidth]{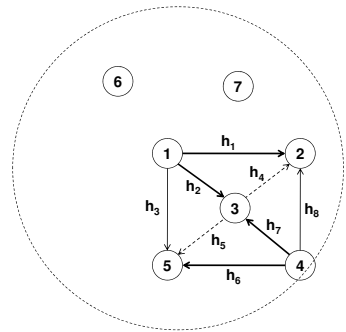}
  \caption{Single cell topology that demonstrates analog network coding and relaying through node $N_3$. The channel gains are denoted with the letter $h$.}
  \label{fig:topology-mac-cc2}
\end{center}
\end{figure}

The topic of concurrent wireless signal transmission jointly employed with network coding is a relatively new research area, while the role of relay has also recently started to be identified as being very critical for the performance of such schemes. For example in~\cite{sagduyu08a} the authors enable ANC at a relay but not for independent users. In~\cite{rimensberger09a} the authors compare different schemes based on ANC with different ML detection techniques. The rate performance of ANC for two-way relaying is analyzed in~\cite{xue07a}. In~\cite{larsson05} the authors introduce a relay topology where the relay encodes the data packets after reception which is similar to digital network coding. In the work presented by Wang and Giannakis in~\cite{giannakis:complex-network-coding} the authors assume that signals from two users are pre-coded before transmitted to a single relay and a single destination. Works that consider the idea of ANC with packets that have been transmitted in the past by a network node were presented in~\cite{popovski06b} with the term bidirectional amplification of throughput (BAT) and in~\cite{katabi07a} with the name ANC. A form of superposition coding in an X topology similar with the topology we have highlighted in Fig.~\ref{fig:topology-mac-cc2} was presented in~\cite{katabi07a}. However, in that work the proposed system attempts to decode independently the overheard and relayed signals leading to higher number of packet failures while the baseline 802.11 MAC is used. Although in all these works the authors study more deeply relaying and ANC, they do not address problems like relay selection in this new context.

When we think about MAC issues in scenarios where ANC is employed, even fewer works exist. One of the most interesting works is the one by Boppana and Shea that proposed the overlapped CSMA protocol~\cite{boppana:overlapped-csma}. The main task of that protocol is to estimate the level of secondary interfering transmissions that another primary transmission can sustain given its perfect knowledge of the signal that intends to cause the interference. This protocol requires significant signaling overhead in order to propagate RTS/CTS messages at least two hops and notify the secondary sender whether it is allowed to proceed or not. Nevertheless, primary and secondary transmissions do not interfere with each other. Also the work by Zhang~\emph{et al.}~\cite{Zhang09a} proposed a similar idea. Very recently the work by Khabbazian~\emph{et al.} presented in~\cite{canc-csail}, proposed the design of a probabilistic MAC based on ANC but only on a theoretical level.

\subsection{Paper Organization}
The rest of this paper is organized as follows. The system model that is used in this paper is presented in Section~\ref{sec:preliminaries}. Subsequently, the distributed channel estimation and information exchange algorithm of our system is analyzed in Section~\ref{sec:channel-estimation}. The mathematical tools for rate estimation under the three possible transmission modes are described in Section~\ref{sec:rate-estimation}. The proposed distributed cooperative MAC protocol and the associated relay selection mechanism are presented in Section~\ref{sec:coopmac}. The signal recovery algorithm is an essential part of our complete system architecture and is described in Section~\ref{sec:collision-recovery}, while Section~\ref{sec:complexity} provides a discussion regarding complexity and implementation issues. In Section~\ref{sec:performance-evaluation} we present comprehensive simulation results for different network traffic patterns. Finally, Section~\ref{sec:conclusions} presents our conclusions and ideas for future work.

\section{System Model and Overview}
\label{sec:preliminaries}
In this paper, we study wireless ad hoc local areas networks where all nodes can be potential relays. Since the proposed protocol optimizes the cooperative transmission for a single hop, this relaxation with respect to the network structure, is possible. Fig.~\ref{fig:topology-mac-cc2} presents a small network that is used throughout this paper for explaining several aspects of the presented algorithms. In this paper we assume that the core of the MAC functionality corresponds to the IEEE 802.11 MAC protocol that operates under the distributed coordination function (DCF)~\cite{IEEE-80211}. Nodes contend for the channel and when the backoff timer expires they use the request-to-send (RTS) and clear-to-send (CTS) floor acquisition mechanism for contacting the intended destination node. This very popular way of randomizing channel access with CSMA/CA ensures that there is only one node that completes successfully the RTS/CTS message exchange and obtains access to the channel. The RTS message is received by relay candidates that indicate their ability to act as relays for the impending transmission with a special message that we describe in later sections. Note that a node may be mobile which means that it might not be able to complete the necessary signalling and thus cannot participate in a cooperative transmission. From previous message exchanges, the relays also collect information about channel estimates in their neighborhood, while they subsequently estimate whether another node can transmit concurrently with the node that just exchanged the RTS/CTS. The aforementioned tasks are accomplished with the \emph{cooperative channel information exchange algorithm} and the \emph{rate estimation algorithm} that are processes that are executed continuously and in parallel with the normal protocol operation. In Fig.~\ref{fig:topology-mac-cc2} for example $N_3$ estimates, according to the latest channel statistics, that $N_4$ can also transmit at the same time with $N_1$, while $N_6,N_7$ might have similar estimates. If interfering transmissions cannot be allowed by any relay, node $N_1$ proceeds with its transmission either cooperatively with the help of $N_3$ (named COOP transmission mode) or directly. Assume now that $N_3$ allows the two transmissions from $N_1$ and $N_4$ to take place concurrently. This task is accomplished with the \emph{cooperative ANC MAC}. Because of the broadcast nature of the channel the two packets/signals will interfere in several physical locations: nodes $N_2$, $N_3$, $N_5$, $N_6$, and $N_7$. In this way, both $N_2$ and $N_5$ have a locally interfered version of the signals that they simply cannot decode. The relay that has been selected with the previous algorithms, forwards its own version of the locally interfered signals to the two destinations. The destinations use then the two versions of the same interfered signals for recovering their respective packet with an \emph{ANC signal decoding algorithm}. The algorithm decodes symbol-by-symbol the interfered packets. This transmission mode is named ANC with overlapped transmissions (ANC-OL). Therefore, our complete system comprises of four algorithms that we describe in the rest of this paper.

\section{Cooperative Channel Information Exchange}
\label{sec:channel-estimation}
It is clear from the introductory description that estimating the channel is necessary both for the decoding algorithm executed at the destinations, but also for the rate estimation. In this paper, channel estimates are obtained after averaging a number of measurements done for each symbol in the preambles and postambles of each control or data packet exchanged at the MAC layer~\cite{argyriou:twc-ancol}. Since the estimation of the channel from preamble/pilot-based schemes is a well known technique~\cite{book:fundamental-wireless}, we do not delve into this topic further. However, for testing if a potential ANC-OL transmission is indeed the optimal choice for transmitting a packet, all the involved channels must be estimated. For example in Fig.~\ref{fig:topology-mac-cc2} all the channel transfer functions shown with the letter $h$ must be estimated in order to be able to test if the specific ANC-OL transmission is efficient (a subset of them in case of COOP). Therefore, a significant number of messages should normally be exchanged even in the simple network of Fig.~\ref{fig:topology-mac-cc2}. In this paper all the necessary channels are estimated by leveraging the transmission of existing control messages in order to avoid additional traffic. The precise algorithm that ensures minimum overhead is shown in Fig.~\ref{fig:distributed-channel-estimation}, while it is described below in detail.

\begin{figure}[t]
\framebox[1.0\linewidth]{
\begin{minipage}[t]{0.9\linewidth}
$est\_channel\_i(pkt)$
\begin{algorithmic}[1]
\IF{$rx\_phy()==RTS$}
\STATE $j=RTS.snd$, $k=RTS.dst$
\STATE $\tilde{h}_{i,j}=estimate\_channel(i,j)$
\IF{$k==i$}
\STATE $\textbf{payload}={\tilde{h}_{i,j}}$
\IF{\textbf{anfl}(i,*,*,*,r)!=0 \& \textbf{anfl}(i',*,*,*,r)!=0}
\STATE $\textbf{payload}=\textbf{payload}+{\tilde{h}_{i',i}}$
\STATE /* This node $i$, and $i'$ have used relay $r$ */
\ENDIF
\STATE $wait(T_{SIFS})$
\STATE $\textbf{dsts}=\{j\}$
\STATE $tx\_phy(CTS,\textbf{dsts},\textbf{payload})$
\ENDIF
\ENDIF
\IF{$rx\_phy()==CTS$}
\STATE $j=CTS.snd$, $k=CTS.dst$
\IF{$k==i$}
\STATE $\tilde{h}_{i,j}=estimate\_channel(i,j)$
\ELSIF{ $i$ is a RELAY}
\STATE $\tilde{h}_{i,j}=estimate\_channel(i,j)$
\ENDIF
\ENDIF
\IF{$rx\_phy()==CTC$}
\STATE $r$=$CTC.snd$, $k_1$=$CTC.dst_1$, $k_2$=$CTC.dst_2$
\IF{$k_1|| k_2 ==i$}
\STATE \textbf{anfl}=\textbf{anfl}+\{$i\rightarrow l_1,k_2\rightarrow l_2,r$\}
\ELSE
\STATE \textbf{anfl}=\textbf{anfl}+\{$k_1\rightarrow l_1,k_2\rightarrow l_2,r$\}
\ENDIF
\ENDIF

\end{algorithmic}
\end{minipage}
}
\caption{Cooperative channel estimation and information exchange algorithm. }%
\label{fig:distributed-channel-estimation}
\end{figure}

\begin{table*}[t]
\caption{Representative organization of the protocol data structures. The X indicates non-zero value.}
\label{tab1}
\centering
\begin{tabular}{| c | c | c | c | c | c | c | c | c | c |}
\hline
\multicolumn{5}{|c|}{$\mathbf{anfl}$}    &\multicolumn{2}{c|}{$\mathbf{channel\_estimates}$} & \multicolumn{3}{|c|}{$\mathbf{rate\_estimates}$} \\
\hline
Src 1 & Dst 1 & Src 2 & Dst 2 & Relay & $S\rightarrow D$ & $S\rightarrow R$,$R\rightarrow D$ & $\widetilde{R}^D$ & $\widetilde{R}^{COOP}$ & $\widetilde{R}^{ANCOL}$ \\
\hline
$N_1$ & $N_2$ &   -    & -     &  - & $\tilde{h}_1$  & - & X &   -   & -\\
$N_1$ & $N_2$ &   -    & -     &  $N_3$ & $\tilde{h}_1$  & $\tilde{h}_2$,$\tilde{h}_4$ & - & X     & -\\
$N_1$ & $N_2$ &   -    & -     &  $N_6$ & $\tilde{h}_1$ & X & - & X     & -\\
$N_1$ & $N_2$ & $N_4$  & $N_5$ &  $N_3$ & $\tilde{h}_1$,$\tilde{h}_6$,$\tilde{h}_3$,$\tilde{h}_8$ & $\tilde{h}_2$,$\tilde{h}_4$,$\tilde{h}_5$,$\tilde{h}_7$ & - & - & X\\%
$N_1$ & $N_2$ & $N_4$  & $N_5$ &   $N_6$ & $\tilde{h}_1$,... & $\tilde{h}_2$,... & -  & - & X\\%
... & & & & & & & & &\\
\hline
\end{tabular}
\end{table*}

Every time a node associates with a specific wireless local-area network (WLAN) this algorithm is initiated while when it diss-associates (also because of mobility) the algorithm stops and the related data structures are cleared. This pseudo-code demonstrates what happens if a control frame/packet is overheard by a node $i$ and how from specific packets we extract information that is useful for channel estimation. The main characteristic of the algorithm is that it leverages the existing RTS/CTS mechanism as many cooperative protocols do~\cite{coopmac,laneman04} and in addition the \emph{clear-to-cooperate} (CTC) message that is introduced in this paper. The precise rules for overhearing and channel estimation are as follows: (1) The first requirement is that all nodes should overhear RTS messages regardless of whether the transmission is intended for them or not and estimate the channel between the transmitting node and themselves (line 3 in the algorithm). (2) All nodes should overhear the CTC message transmissions of their neighbors. Each node should maintain a data structure that it should contain the nodes and the associated relay that were involved in an overheard COOP or ANC-OL transmission. This data structure is named $\mathbf{anfl}$ in the algorithm and its organization can be seen in Table~\ref{tab1}. Furthermore, this data structure should be updated continuously with more recent information that is extracted from overheard CTC messages (lines 23-30 in the algorithm), and its size should reflect the node resources. This information will be used for identifying the specific channels/nodes that can be part of a COOP or ANC-OL transmission. To understand how this works consider the example in Fig.~\ref{fig:topology-mac-cc2}. In this figure $N_2$ overhears cooperative transmissions (the CTC message) between $N_4$ and $N_5$ with $N_3$ being the relay. In a symmetrical fashion, $N_5$ overhears the cooperative transmission from $N_1$ to $N_2$ with the help of $N_3$. (3) A node should piggyback in its outgoing CTS message the results of the channel estimation only for channels that are formed between another node and themselves, but only if both have used the same relay in the past. This check is performed in lines 6-9 of the algorithm with information that is extracted from the $\mathbf{anfl}$ data structure that contains monitored data from several past relayed transmissions. To continue our previous example when $N_2$ sends a CTS for responding to an RTS from $N_1$, it includes in the CTS response not only the estimate $\tilde{h}_1$, but also the estimate that it has for $\tilde{h}_8$ which was obtained from previous transmissions of RTS messages from $N_4$ (Recall a few lines above that $N_4$ and $N_3$ were included in the $\mathbf{anfl}$ data structure of $N_2$). One way to summarize this functionality is that in this way a relay can obtain the information for channels that it cannot directly estimate ($\tilde{h}_8$ and $\tilde{h}_3$ here).

This adaptive flow monitoring technique with the $\mathbf{anfl}$ data structure, increases the channel information at the relay only when it could be needed. Also it is important to note that from an implementation perspective, when a tagged node experiences at the MAC layer a diss-association from another node, then the channel estimates that involve the disconnected node, are removed from the local memory and also the entries in the $\mathbf{anfl}$ data structure that involve this node.

\section{Rate Estimation of Cooperative and Interfering Transmissions}
\label{sec:rate-estimation}
The next question is as follows: How does the system select which secondary/interfering transmission is optimal? Naturally, a secondary transmission should be selected to interfere \textit{iff} the ANC-OL mode will increase the sum-rate not only when compared to the direct transmission, but also when compared to a COOP transmission that employs amplify-and-forward (AF)~\cite{laneman04,coopmac}. To do so it must be evaluated analytically, and more importantly during run-time, which type of cooperation is the most efficient. The only issue is that this decision can only be made by the relay since it is the only node in the network configuration that can obtain all the necessary information for doing so as we described in Section~\ref{sec:channel-estimation}.

In the general case of cooperative systems, the transmitter may select to use cooperative transmission when a desired rate is not met with a direct transmission. However, without loosing generality we assume that with the proposed protocol the optimal mode is always selected whether it is ANC-OL, COOP, or Direct. Now consider that the channel bandwidth is $W$, the transmitter power $P$, additive white Gaussian noise (AWGN) with zero mean and variance $\sigma^2$, and $\gamma_i=|h_i|^2$. If we assume Rayleigh block fading channels where the attenuation is considered constant throughout the transmission of a single frame then the SNR between two nodes in our system is given by $SNR=\frac{P\gamma}{\sigma^2}$. The estimated rate of the Direct transmission mode is then:
\begin{equation}
\label{ineq:1}
\widetilde{R}_{DIR}=W\cdot log_2(1+\frac{P\tilde{\gamma}_1}{\sigma^2}).
\end{equation}
On the other hand, the estimated rate of the cooperative transmission COOP that occurs in two orthogonal time slots for the example in Fig.~\ref{fig:topology-mac-cc2} will be~\cite{laneman04}:
\begin{eqnarray}
\label{ineq:2}
\widetilde{R}_{COOP}&=&\frac{W}{2} \cdot min\big\{log_2(1+\frac{P\tilde{\gamma}_2}{\sigma^2}),\\
&&log_2(1+\frac{P\tilde{\gamma}_1}{\sigma^2}\frac{P\tilde{\gamma}_2\tilde{\gamma}_4g^2}{\sigma^2(1+\tilde{\gamma}_4g^2)})\big\}. \nonumber
\end{eqnarray}
If we consider the overhead of the complete protocol we design in the next section, the cooperative scheme will be more efficient when it is
\begin{eqnarray}
\label{ineq:3}
&&\frac{L}{\widetilde{R}_{COOP}}+T_{OVHD,COOP}<\frac{L}{\widetilde{R}_{DIR}}.
\end{eqnarray}
The aforementioned condition can also be interpreted as follows: The COOP transmission mode is more efficient when the time duration of the cooperative transmission is shorter from the direct transmission based on the estimated rate, plus the associated protocol overhead ($T_{OVHD}$) that is incurred by the cooperative protocol. Similar conditions are used by other cooperative protocols~\cite{coopmac}. This condition can also determine the optimal packet length $L^*$ for which direct or cooperative transmission is optimal.

Now we present the estimated sum-rate of the ANC-OL transmission from the present relay and for the unicast transmissions depicted in Fig.~\ref{fig:topology-mac-cc2}, i.e.  $N_1 \rightarrow N_2$ and $N_4 \rightarrow N_5$. This sum-rate expression for two interfering transmissions incorporates the overheard information that is used for decoding the respective signals/packets at each receiver. This will be equal to~\cite{argyriou:twc-ancol}:
\begin{eqnarray}
\widetilde{R}_{ANCOL}&=& W \cdot\log_2 \Big (1+\frac{P\gamma_1}{\sigma^2}+\frac{P\gamma_8}{\sigma^2}+\frac{P\gamma_2\gamma_4g^2}{\sigma^2(1+\gamma_4g^2)}\nonumber\\
&+&\frac{P\gamma_4\gamma_7g^2}{\sigma^2(1+\gamma_4g^2)}+\frac{P^2\gamma_1\gamma_4\gamma_7g^2}{\sigma^4(1+\gamma_4g^2)}\\
& +&\frac{P^2\gamma_2\gamma_4\gamma_8g^2}{\sigma^4(1+\gamma_4g^2)}\nonumber-\frac{P^2\gamma_4 Re(h_1h_2^*h_7h_8^{*})g^2}{\sigma^4(1+\gamma_4g^2)} \Big ).  \label{ria}
\end{eqnarray}
The above formula is not a pre-requisite for the operation of the proposed rate estimation algorithm and of course the entire protocol. Similar transmission modes like ANC-OL could be utilized in conjunction with a suitable analytical rate expression (E.g.~\cite{katabi07a}). Also for the ANC-OL mode to be more efficient than COOP in addition to inequality~\eqref{ineq:3}, the following condition must be true:
\begin{eqnarray}
\label{ineq:4}
\frac{L}{\widetilde{R}_{ANCOL}}+T_{OVHD,ANCOL}<\frac{L}{\widetilde{R}_{COOP}}+T_{OVHD,COOP}.
\end{eqnarray}
Relays use the previous rate estimation expressions for estimating the possible rate between for all the available channel estimates that they have stored for their neighbors. These results populate a data structure like the one depicted in Table~\ref{tab1}, and in this case we name it $\mathbf{rate\_estimates}$. As we will see in the next section, a relay decides if it will notify another node regarding its ability to cooperate with the use of busy tones. A busy tone is a narrowband signal transmitted at the maximum allowed power of the wireless standard. This is accomplished on-demand, i.e. when another node desires to transmit.

\section{Cooperative ANC MAC (CANC-MAC)}
\label{sec:coopmac}
The two previous algorithms for cooperative channel information exchange and rate estimation are essential for the operation of our system but they do not affect directly the channel access mechanism. Now we describe the third central component of the complete system that is the cooperative analog network coding MAC (CANC-MAC) protocol. The proposed protocol does not affect the contention and channel access mechanism but only the cooperative packet transmission procedure. It is important to be clear that the adoption of the well-known and understood binary exponential backoff algorithm allows one node to obtain access to the channel at a specific time instant and transmit an RTS/CTS. Therefore, it is impossible for two nodes to successfully complete the RTS/CTS exchange. Since the two nodes that are about to be involved in a communication are identified with the method above, the problem that remains to be addressed is to identify which node can be the optimal relay and if there are any additional nodes that can transmit concurrently.

\begin{figure}[t]
\framebox[1.0\linewidth]{
\begin{minipage}[t]{0.9\linewidth}
$tx\_data(\mathbf{D},\mathbf{payload})$
\begin{algorithmic}[1]
\STATE $execute\_backoff()$
\STATE $\textbf{dsts}=\{D\}$
\STATE $tx\_phy(RTS,\textbf{dsts},\textbf{payload})$, $wait(T_{SIFS})$
\IF{$rx\_phy()==CTS$}
\STATE $wait(T_{SIFS})$,$check\_channel(T_s)$
\IF{busy tone received?}
\STATE $\textbf{dsts}=\{Relay^{ANCOL}_{opt},D\}$
\ELSE
\STATE $wait(T_s)$, $check\_channel(T_s)$
\IF{busy tone received?}
\STATE $\textbf{dsts}=\{Relay^{COOP}_{opt},D\}$
\ENDIF
\ENDIF
\ENDIF
\FOR{all slots until N}
\STATE $check\_channel(T_s)$
\IF{$rx\_phy()==CTS||CTC$}
\STATE $wait(T_{SIFS})$, $tx\_phy(DATA,\textbf{dsts},\textbf{payload})$
\ENDIF
\ENDFOR
\end{algorithmic}
$relay\_overhear(S,D)$
\begin{algorithmic}[1]
\STATE update $\mathbf{rate\_estimates}$, $\mathbf{channel\_estimates}$
\STATE $wait(T_{SIFS})$
\IF{($\widetilde{R}^{COOP}>\widetilde{R}^{DIR}$)}
\IF{($\widetilde{R}^{ANC}>\widetilde{R}^{COOP}$)}
\STATE  $tx\_phy(busy\_tone)$
\ELSE
\STATE $wait(T_s)$, $tx\_phy(busy\_tone)$
\ENDIF
\STATE $relay\_backoff(\tilde{R},N)$, $tx\_phy(CTC,\textbf{dsts})$
\ENDIF
\end{algorithmic}
\end{minipage}
}
\caption{Pseudo-code of the main functionality of the proposed cooperative ANC MAC protocol at the sender and the relay. }%
\label{fig:canc-mac}
\end{figure}

\subsection{Basic Protocol and Busy Tones}
The $tx\_data()$ subroutine in the pseudo-algorithm of Fig.~\ref{fig:canc-mac} depicts the actions executed at a sender when it desires to transmit a data packet. Let us assume that an RTS/CTS message exchange has finished (line 5 in the previous subroutine) and several relays have updated the $\mathbf{rate\_estimates}$ as we explained in the previous section.  Then the potential relays indicate their ability to relay a transmission by using busy tones that are transmitted after a time duration equal to $T_{SIFS}$ after the end of the CTS transmission\footnote{For being compatible with the basic RTS/CTS message exchange of existing devices the transmission of the busy tone should be delayed for the duration of one more slot. This will allow a legacy node to start transmitting a data frame before any relay indicates its intention with busy tones (see Fig.~\ref{fig:relay-prioritization}).}. Note that busy tones are also transmitted in the same channel while there is no separate control channel. The conditions for transmitting busy tones are the following: A busy tone is transmitted from a relay candidate in the first slot after $T_{SIFS}$, if the relay desires to indicate that the ANC-OL mode is efficient for improving the rate of the system by combining the indicated transmission with another transmission. This is indicated in line 5 of the $relay\_overhear()$ subroutine in Fig.~\ref{fig:canc-mac}. When no busy tone is transmitted after $T_{SIFS}$ plus $T_s$, this means that this transmission cannot use the ANC-OL mode jointly with another transmission based on the latest estimate by the relay(s). On the other hand, the first slot after CTS plus $T_{SIFS}$ remains idle, and a busy tone is transmitted by a relay in the second slot, when the rate can be improved by enabling the COOP mode (again depicted in the $relay\_overhear()$ subroutine in Fig.~\ref{fig:canc-mac}). Similarly with before, several potential relays can transmit a busy tone. The optimal one has again to be selected in a similar way as in the case of ANC-OL.

Finally, if no busy tone is transmitted in any of the first two slots after $T_{SIFS}$, the Direct transmission mode is selected instead. In this last case, the node that obtained the channel and sent the first RTS will send directly the data packet waiting at most $T_{SIFS}$ plus $2T_s$ after the CTS reception. This minor delay of two time slots is very short when compared to the overall performance benefits of the proposed scheme. Note that busy tones are used since other relay candidates might also transmit a busy tone in the same slot (e.g. nodes $N_6,N_7$), which means that at least one node can be used for ANC-OL.

\subsection{Relay Prioritization}
\begin{figure}[t]
\begin{center}
  \includegraphics[keepaspectratio,width = 0.99\linewidth]{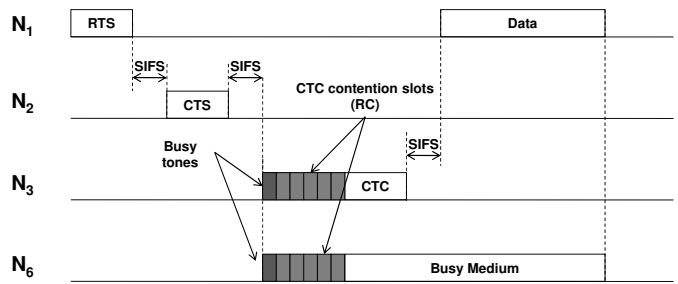}
  \caption{Message exchange for optimal relay selection through prioritization.}
  \label{fig:relay-prioritization}
\end{center}
\end{figure}

The next question is the following: How does the system treat multiple relay candidates? From all the potential relay nodes, the one with the highest possible increase in the transmission rate should obtain the channel and be used as a relay. To solve this problem a separate round is introduced during which relays are allowed to contend for this role. Fig.~\ref{fig:relay-prioritization} presents how two relays contend for the relaying opportunity. We named this process \emph{the relay contention round} and it works as follows. After the relay nodes transmit their respective busy tones, they set the value of a special backoff counter. The contention slot counter at a relay is set in terms of slots as $T_{RBKF}=(2\cdot N - \lfloor \widetilde{R} \cdot N \rfloor ) \cdot T_s$, where $N$ is the maximum value for the contention slots. The value of $N$ depends on the maximum allowed delay and it should be configured for the complete network during the initialization phase. What this formula does is that it allocates a smaller number of slots for nodes that can achieve the higher rate with any transmission mode\footnote{Note that $\widetilde{R}$ is the normalized estimated rate gain from any transmission mode and takes values between 1 and 2, with 2 denoting the maximum gain, i.e. two packets/slot.}. In this way the relay with the highest possible rate obtains the channel by minimizing the number of slots it has to wait before it transmits a CTC message. Other potential relays that overhear a transmitted CTC, can infer safely that another more optimal node will relay the impending transmission, and they simply stop the $T_{RBKF}$ timer. Now, the overhead in time slots that the proposed protocol introduces can now be easily derived from Fig.~\ref{fig:relay-prioritization} as follows:
\begin{equation}
T_{OVHD}=T_{RTS}+2T_{CTS}+3T_{SIFS}+2T_{s}+T_{RBKF}.
\end{equation}
After the $T_{RBKF}$ timer expires, the relay transmits a CTC message towards both nodes that should transmit concurrently (line 9 in the $relay\_overhear()$ subroutine of Fig.~\ref{fig:canc-mac}). CTC is essentially a CTS message that contains two destination addresses and indicates to the senders that the concurrent transmission can take place after $T_{SIFS}$ allowing thus a synchronized collision. From the perspective of the initial sender of the RTS, the process that checks the existence of CTC and the transmition the actual data packet is handled in lines 15-20 of the $tx\_data()$ subroutine in Fig.~\ref{fig:canc-mac}. The main advantage of the proposed protocol is that the receivers do not need to explicitly identify the ANC-OL transmission since they know that signals that are received after the CTC will interfere. The only need by the receiving nodes is to check the CTC header and make sure that they are one of the intended destinations of the impending ANC-OL. This means that they can employ the signal recovery algorithm that we describe in the next section directly after the reception of the interfered packets.

\section{Recovery of Interfered Packets}
\label{sec:collision-recovery}
At this stage we have reached the point in the overall system functionality where an ANC-OL transmission has been completed. Now the two interfered signals will need to be jointly decoded. For simplifying the notation and the explanation of the algorithm here, we use again the example in Fig.~\ref{fig:topology-mac-cc2} to demonstrate this process. Let us denote the transmitted packets/signals as $x_A$ and $x_B$. The transmitted signals in this example originate from $N_1$ and $N_4$ and are received by the intended receivers $N_2$ and $N_5$ respectively, and also by the relay $N_3$. With the ANC-OL mode the relay $N_3$ will broadcast the locally interfered version of the two signals. For recovering the two independent packets, we apply joint decoding of the locally interfered and forwarded versions of the interfered packets. Before this process takes place, the receiver identifies the packet preamble that is contained in each version of the two aforementioned signals and then it aligns them at the symbol-level~\cite{argyriou:twc-ancol,argyriou2010coml}. Subsequently, a maximum-likelihood (ML) detector is applied for the symbols that belong to the two different packets.

For expressing mathematically our algorithm let us denote with $\mathcal{X_A},\mathcal{X_B}$ the  fixed symbol dictionaries that depend on the modulation scheme that the two senders use. Let also $P$ denote the power allocated at each sending node, while $g$ is the power allocation factor at the relay. Finally let also $n$ denote the noise at the corresponding receiver that is a circularly complex Gaussian random variable, i.e. $\mathcal{CN}(0,1)$. From Fig.~\ref{fig:topology-mac-cc2} we can see that the direct signal that will be received at the destination $N_2$ is
\begin{equation}\label{ML_detection_joint}
y_{N_2}=\sqrt{P}h_1x_A+\sqrt{P}h_8x_B+n_{N_2},
\end{equation}
while the forwarded signal from the relay is
\begin{equation}\label{ML_detection_joint}
y_{N_3,N_2}=\sqrt{P}h_2h_4gx_A+\sqrt{P}h_7h_4gx_B+h_4gn_{N_3}+n_{N_2}.
\end{equation}
If node $N_2$ combines the direct and relayed signals with a single ML demodulation step, the estimation will take the form
\begin{eqnarray}\label{ML_detection_joint}
&&    (\tilde{x}_A,\tilde{x}_B)_{N_2} = \arg \min_{x_A \in \mathcal{X_A},x_B
\in \mathcal{X_B}} \Big \{ \| y_{N_2}-\sqrt{P}h_1x_A\nonumber\\ %
        &-&\sqrt{P}h_8x_B\| + \| y_{N_3,N_2}-\sqrt{P}h_2h_4gx_A\nonumber\\
        &-&\sqrt{P}h_4h_7gx_B \| \Big \}.%
\end{eqnarray}
At the second receiver, a similar signal recovery formula can be written. The parameters $\sqrt{P} h_4h_7g$, and $\sqrt{P}h_2h_4g$ are obtained by using the training symbols that are inserted in the preambles of RTS, CTS, and CTC packets as we explained in detail earlier in this paper.  The performance of the algorithm that we just described and is summarized in~\eqref{ML_detection_joint}, was studied in~\cite{argyriou:twc-ancol} where as we said in the Introduction we also developed a thorough sum-rate analysis for the case of two independent senders.

\section{Complexity and Implementation Issues}
\label{sec:complexity}
As we pointed out in the Introduction, the proposed system consists of different algorithms of varying complexity. Therefore, we think it is necessary to provide a brief discussion regarding the complexity and implementation issues that might arise.

A general characteristic of our system is that relaying decisions are only made for a single hop since the algorithms operate at the link layer. A node that decodes with the proposed algorithm two interfered packets successfully, it will transmit this packet to its next hop by following the same process. Therefore, in the case that the network has multiple hops, the additional nodes perform the same tasks again but only for their own neighborhood, limiting thus the number of nodes that have to be considered in our algorithms.

Another issue is that several of the algorithms are executed at the relays which might introduce significant overhead. However, we do not expect that this is the case for the following reasons. First, the channel estimation is usually a process applied in existing WLANs while the relay only has to overhear RTS/CTS messages for performing this task. Second, in existing WLAN devices rate selection algorithms are also applied and are primarily vendor-specific. Third, the rate estimation algorithm described in Section~\ref{sec:rate-estimation} requires only a few numerical calculations. Therefore, current hardware is capable of supporting these algorithms. Finally, we should also note that in our network every node is a potential relay since we assume this a collaborative  network and nodes are willing to share their resources for maximizing the total throughput. Of course if the local resources are not sufficient, then a node can refrain from being used as a relay.

Regarding the complexity of the proposed decoding scheme is essentially a V-BLAST~\cite{foschini96} decoder that is generally characterized by exponential computational complexity in both the number of transmitters and the size of the symbol constellation. But since in this case the number of transmitters is two, the decoding complexity is similar to a 2x2 MIMO system~\cite{book:fundamental-wireless}. There are other suboptimal lower complexity detection methods for V-BLAST such as zero-forcing (ZF) detection or minimum mean squared error (MMSE) detection. However, these methods result in significant performance degradation when compared to ML detection. We expect that this is the only algorithm that needs new hardware processing functionality since it requires different signal processing algorithm at the PHY.

\section{Performance Evaluation}
\label{sec:performance-evaluation}
The performance of the proposed system is evaluated through computer simulation. We assume that nodes are randomly placed in a single cell and that pairs of backlogged nodes communicate to each other. We implemented CANC-MAC and we evaluated the performance in terms of MAC layer throughput (including the overheads) and packet transmission delay under different channel conditions. All nodes are assumed to be backlogged with traffic while results are obtained for 10,000 packet transmissions. The channel access timing parameters are similar with 802.11 ($T_{SIFS}$=16$\mu sec$, $T_{DIFS}$=34$\mu sec$). Regarding the lower layer parameters we assume a channel bandwidth of $W=20$ MHz, while the same Rayleigh fading path loss model was used for all the channels. Our assumptions in this case include a frequency-flat fading wireless link that remains invariant per transmitted PHY frame, but may vary between simulated frames. The channel quality is captured by the average received SNR $\gamma$ of the wireless link. Since the channel varies from frame to frame, the Nakagami-$\eta$ fading model is adopted for describing $\gamma$~\cite{book:fundamental-wireless}. This means that the received SNR per frame is a random variable, where we assume $\eta=1$ for Rayleigh fading. The noise over the wireless spectrum is additive white Gaussian noise (AWGN) with the variance of the noise to be $10^{-9}$ at every node/link. Regarding specifics of CANC-MAC, the number of \emph{different} ANC-OL and COOP transmissions that are monitored and kept in the data structure was 20 while the maximum number of backoff slots in the relay contention round was set to $N=10$. For comparing our protocol, we also implemented a typical relaying scheme named COOP-MAC, that employs orthogonal cooperative transmissions without interfering signals~\cite{laneman04}.

Finally, we investigated the impact of traffic pattern changes. For the ANC-OL mode, a change in the next hop of one of the unicast transmissions  will affect the performance of the channel estimation and ML detection algorithms since they have to be executed for a different next hop destination. To this aim we devised \textit{Scenario 1} where a source-destination pair is constant throughout the simulation, and \textit{Scenario 2} where nodes were alternating their next-hop destination node after the transmission of 500 consecutive packets. This last scenario is one way to simulate the behavior of nodes that act as routers in multi-hop or mobile communication scenarios.

\subsection{Throughput vs. Number of Nodes}
\begin{figure}[t]
\begin{center}
\includegraphics[keepaspectratio,width = 0.5\linewidth]{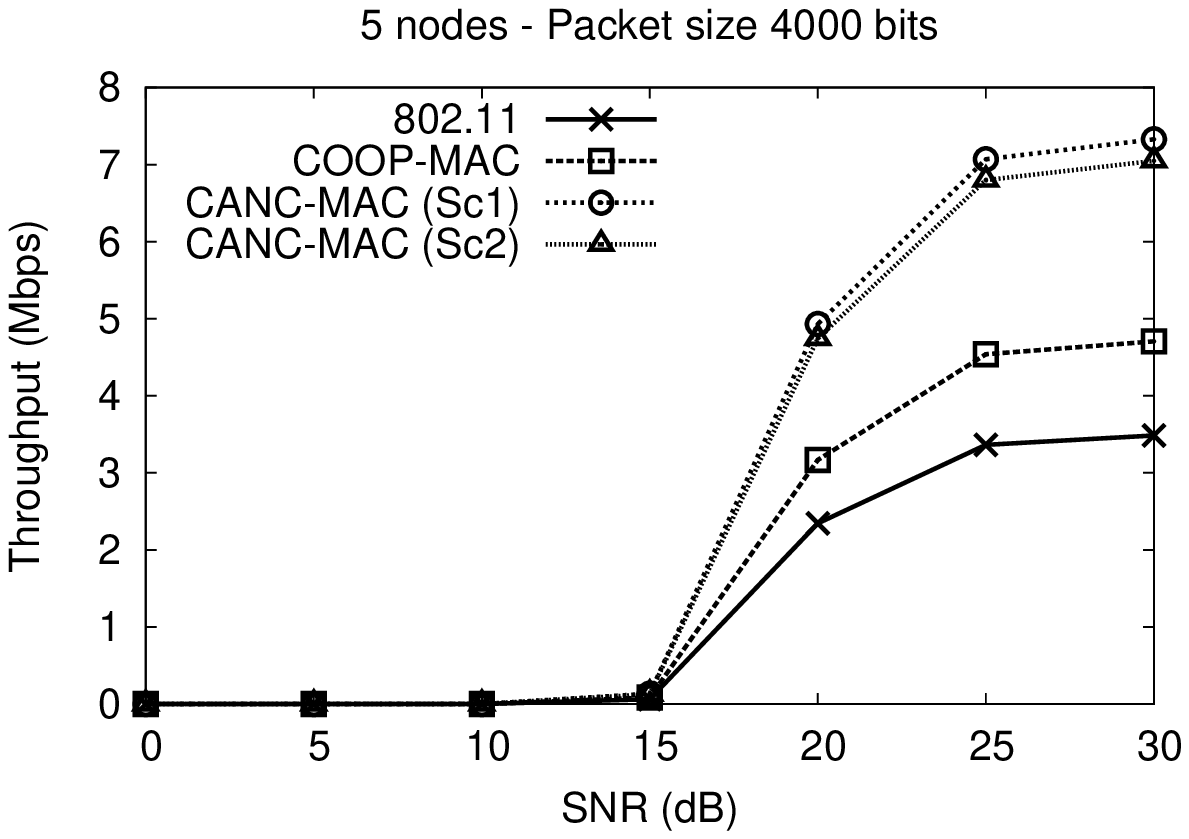}\hspace{-0.2cm}%
\includegraphics[keepaspectratio,width = 0.5\linewidth]{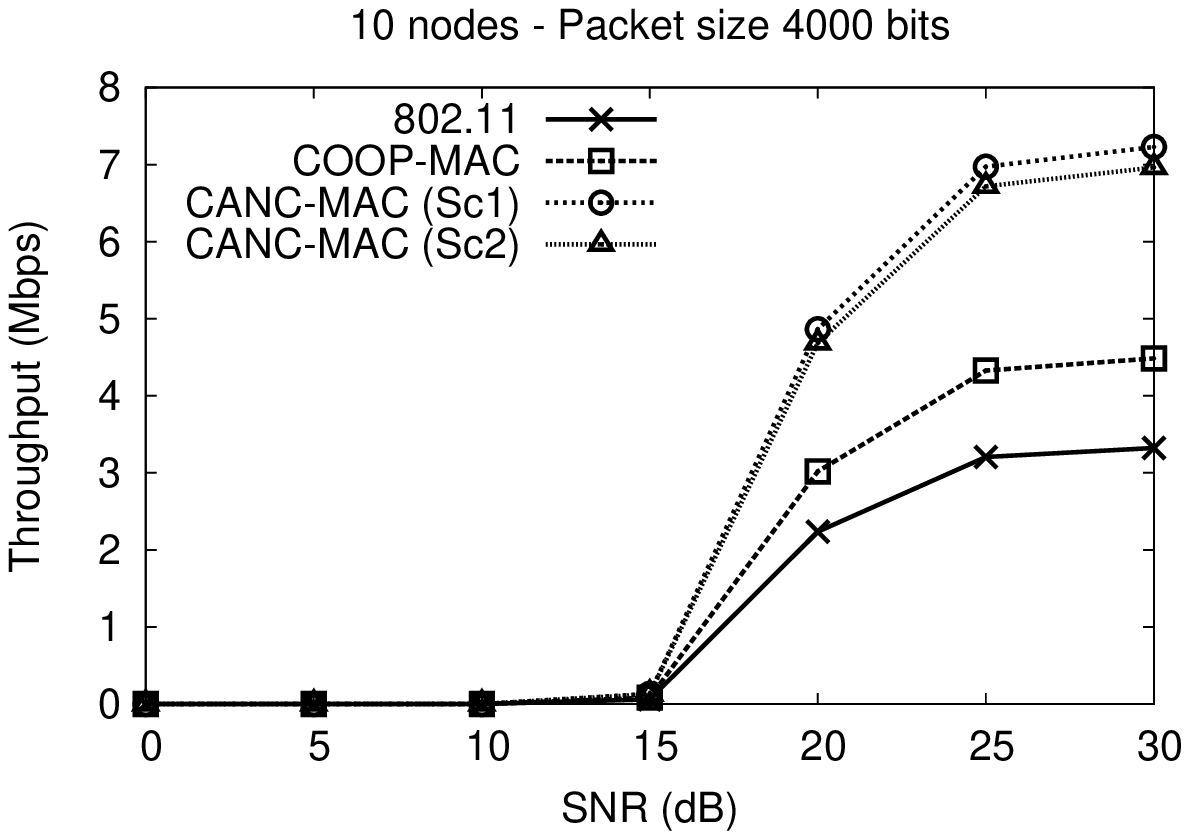}
\includegraphics[keepaspectratio,width = 0.5\linewidth]{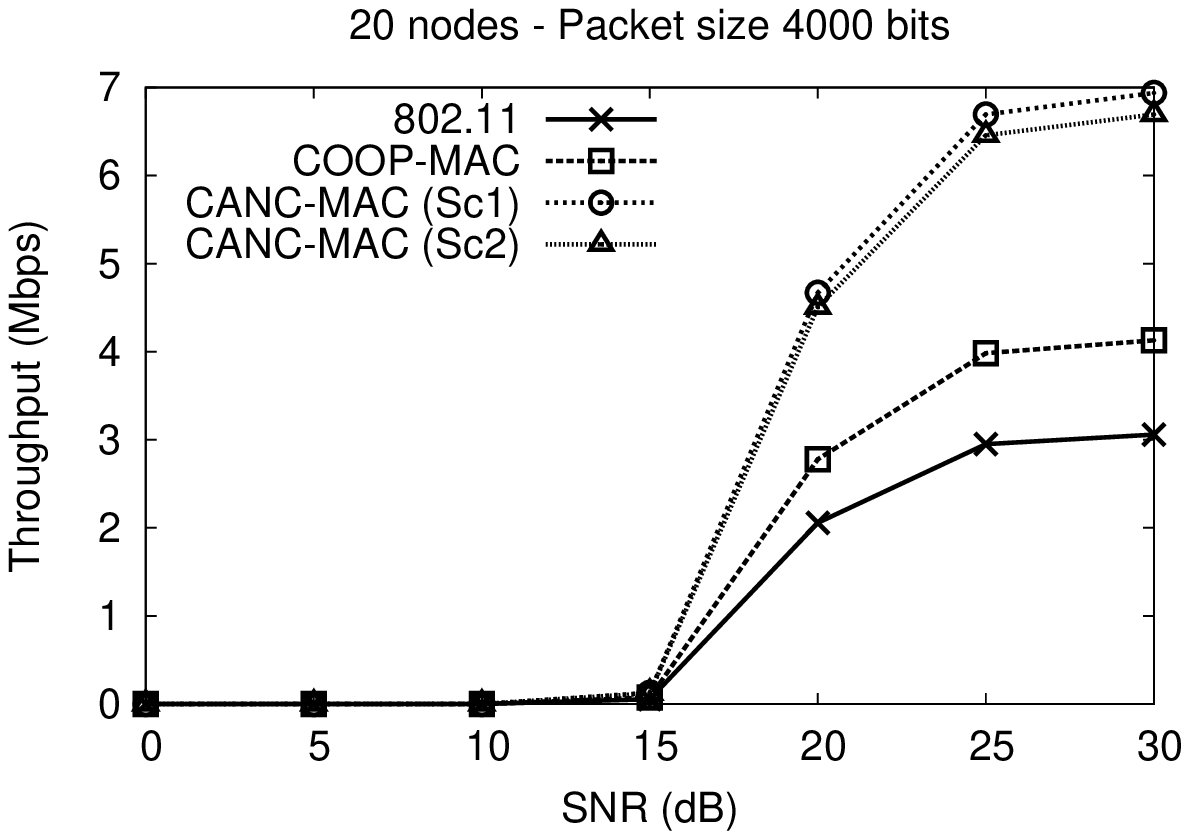}\hspace{-0.2cm}%
\includegraphics[keepaspectratio,width = 0.5\linewidth]{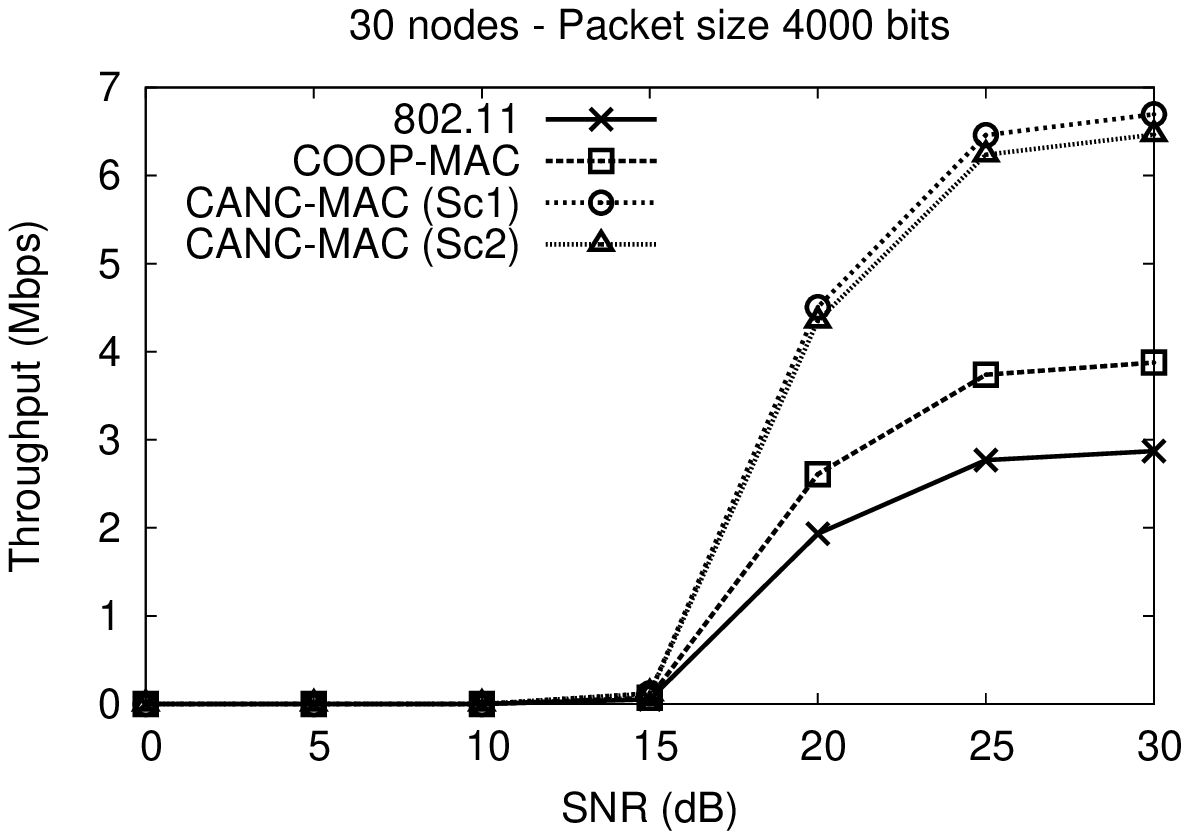}
 \caption{Simulation results of the aggregate network throughput for different channel conditions and different number of nodes. Packet size of 4000 bits is used.}
  \label{fig:anc-throughput1}
\end{center}
\end{figure}

In Fig.~\ref{fig:anc-throughput1} we present the aggregate MAC layer throughput results in the complete network for different number of nodes and for different SNR of the wireless channel. The last parameter is important to be evaluated since it affects the performance of the ML detector that is executed at the receivers. The results are very representative of the performance of complete system we propose since they show that for a higher number of nodes the aggregate MAC layer throughput can remain very high. Therefore, the impact of having a high rate of enforced interfering transmissions when the number of nodes is increased, is mitigated by the proposed cooperative protocol and the associated signal recovery algorithm. It is also interesting to note that for the traffic \emph{Scenario 2} (Sc2) the performance of the proposed scheme is barely impacted by the more frequent changes in the traffic flow. The number of nodes seems to have only minor impact in the performance of the CANC-MAC in \emph{Scenario 2} when compared to \emph{Scenario 1}. The reason for this performance difference is that as the number of nodes that contend for the channel is increased, the time period between two successive packet transmissions takes longer. This fact increases the time duration until the channel information exchange and estimation algorithm updates the available information of a node.

It is important to understand that with the proposed CANC-MAC the performance is always lower-bounded by the baseline COOP-MAC which means that it cannot become worse both theoretically but also practically. One way to explain this intuitively is to think that for low SNR the performance of ML detection is naturally not very good which in practice means that ANC is not used frequently. However, even with the baseline 802.11 or COOP-MAC, the performance is also poor because of the higher bit error rate (BER) of every link. Therefore, CANC-MAC works well and in pace with the performance that we would expect from IEEE 802.11 and COOP-MAC.

\subsection{Throughput vs. Payload Size}
\begin{figure}[t]
\begin{center}
\includegraphics[keepaspectratio,width = 0.5\linewidth]{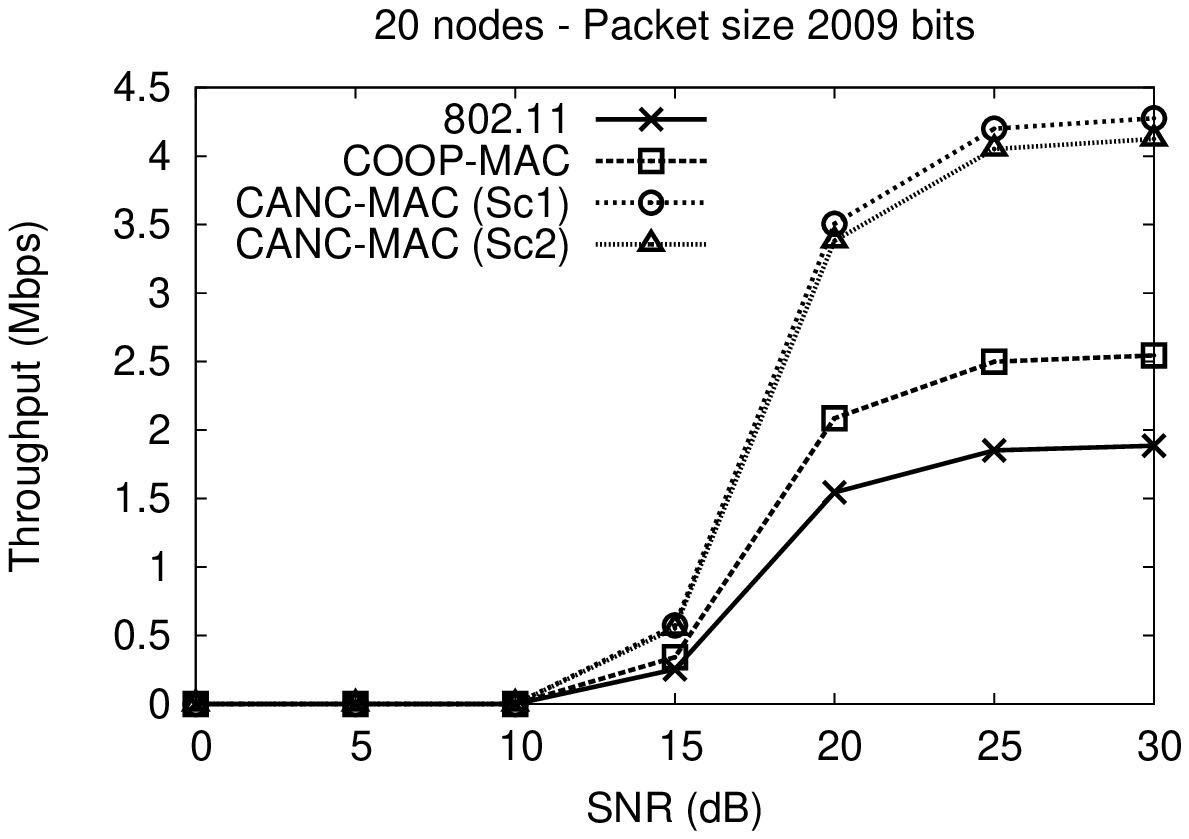}\hspace{-0.2cm}%
\includegraphics[keepaspectratio,width = 0.5\linewidth]{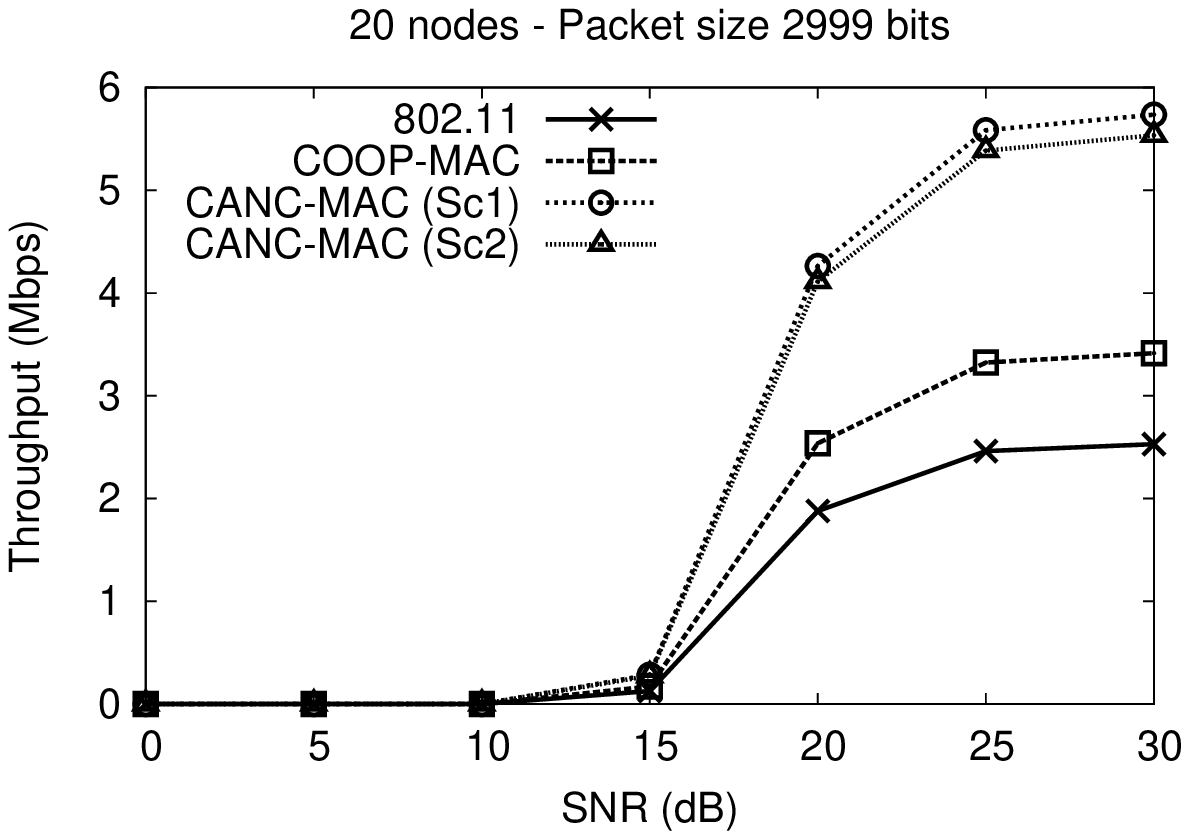}
\caption{Simulation results of the aggregate network throughput for different channel conditions and different packet sizes.}
  \label{fig:anc-throughput2}
\end{center}
\end{figure}

Next we evaluate the performance of CANC-MAC for different payload sizes. Results for packet sizes of nearly 2000 and 3000 bits can be seen in Fig.~\ref{fig:anc-throughput2}. The results are consistent with our previous results for a packet size of 4000 bits, although the aggregate throughput is lower because of the smaller packet size. It is important to note that for higher payload size, the performance increase of CANC-MAC over COOP-MAC is becoming more important. The reason is that the efficiency of the ANC-OL mode is translated to two successful packet transmissions which means higher performance gain from a single interfered transmission. Furthermore, the reduced number of contention rounds that a node has to go through results in an additional improvement of the information rate besides the fact that two packets are transmitted in one slot. Also it is important to see in this figure that in the lower SNR regime the performance of all the protocols is improved as the packet size becomes smaller. However, for a larger packet size the SNR regime under which any protocol improves its performance needs to become substantially higher. For example for a packet size of 3000 bits or higher, a channel SNR of 15dB is needed in order to start observing a meaningful network throughput.

The same observation also holds for \emph{Scenario 2}. We see that in general the impact of packet size variations, or the number of nodes in the previous subsection, have no impact and minor impact on the performance respectively. The performance reduction is purely from the overhead of having to stop the ANC-OL mode to the next hop, and then complete two successful unicast packet transmissions in order to identify new candidates for ANC-OL. However, we believe that even \emph{Scenario 2} is unlikely to happen in reality since the frequent changes in traffic pattern will only probably happen in scenarios of high mobility.

\subsection{Packet Transmission Delay vs. Number of Nodes}
\begin{figure}[t]
\begin{center}
\includegraphics[keepaspectratio,width = 1.0\linewidth]{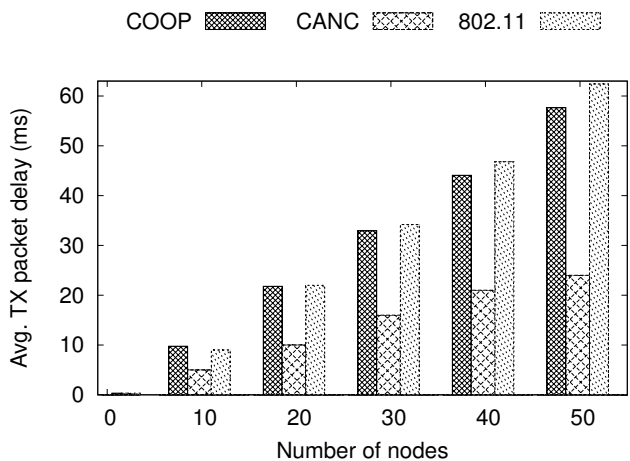}\hspace{-0.2cm}%
 \caption{Channel access delay for the three protocols under test.}
  \label{fig:anc-delay-vs-nodes}
\end{center}
\end{figure}

Results for the packet transmission delay versus the number of nodes can be seen in Fig.~\ref{fig:anc-delay-vs-nodes}. Regarding the performance of the COOP-MAC protocol it reduces the delay when compared to IEEE 802.11 but only because it reduces the number of re-transmissions. The lower BER corresponds to lower packet error rate (PER) and eventually to a reduced number of retransmissions. On the contrary CANC-MAC combines the benefit that diversity provides in combination with the use of cooperative decoding, and also the benefit of transmitting two units of information in a single time slot. In our results in Fig.~\ref{fig:anc-delay-vs-nodes} the additional benefit of CANC-MAC over COOP-MAC is obvious but the delay is not exactly reduced by half as we would expect. Also note that as the number of nodes is increased with CANC-MAC, the rate at which the delay is increased has similar trend with the other two protocols. The explanation for these results is provided below. With the ANC-OL mode a single packet is experiencing a higher transmission delay since it takes slightly longer to access the channel because of the altered protocol procedure. This is because the proposed protocol introduces an overhead even for the transmission of a single packet. However, if the average service time for each packet is considered, then the total delay for each packet is lower with CANC-MAC since it is serviced faster from the transmission queue. When a node sends an RTS before the data packet, our protocol is indirectly "fishing" for another suitable packet that could be transmitted from the HOL position in the queue of another node. Therefore, the average transmission time of packets in the complete network is theoretically reduced by half for fully backlogged nodes and without any protocol overhead. Of course in the case that nodes do not have packets to transmit, we expect that performance gains will be reduced.

\section{Conclusions}
\label{sec:conclusions}
In this paper we presented a cooperative MAC protocol that pro-actively enforces packets to interfere in distributed wireless local area networks. The protocol ensures that when two nodes desire to transmit packets to independent destinations, they coordinate with minimal overhead with a third relay node for concurrently transmitting over the wireless channel. The relay is responsible for ensuring that the desired packets can be decoded and recovered at the respective destinations by using analytical rate expressions. To enable distributed uncoordinated operation of the protocol, we introduce a relay selection mechanism so that the optimal relay can be selected in terms of its ability to increase the achieved transmission rate. Performance results showed the efficacy of our proposed scheme in terms of both throughput and delay. In our future work we plan first to investigate in more detail the necessary protocol enhancements in multi-hop scenarios where more than two transmissions may interfere.


\begin{biography}%
{Antonios Argyriou} received the Diploma in electrical and computer engineering from Democritus University of Thrace, Greece, in 2001, and the M.S. and Ph.D. degrees in electrical and computer engineering as a Fulbright scholar from the Georgia Institute of Technology, Atlanta, USA, in 2003 and 2005, respectively.

Currently, he is a tenure-track faculty member at the Department of Computer and Communications Engineering, University of Thessaly, Greece. From 2007 until 2010 he was a Senior Research Scientist at Philips Research, Eindhoven, The Netherlands. From 2004 until 2005, he was a Senior Engineer at Soft.Networks, Atlanta, GA. Dr. Argyriou currently serves in the editorial board of the \textit{Journal of Communications}. He has also served as guest editor for the \textit{IEEE Transactions on Multimedia} Special Issue on Quality-Driven Cross-Layer Design, and he was also a lead guest editor for the \textit{Journal of Communications}, Special Issue on Network Coding and Applications. Dr. Argyriou serves in the TPC of several international conferences and workshops in the area of communications, networking, and signal processing. His current research interests are in the areas of communication systems and computer networks. He is a member of IEEE.
\end{biography}

\end{document}